\documentclass[aps,prl,showpacs,showkeys,amsmath,amssymb,
twocolumn,
floatfix,
superscriptaddress
]{revtex4}

\usepackage{amsmath}
\usepackage{graphicx}
\usepackage{multirow}
\usepackage{longtable}
\usepackage{rotating}

\usepackage{amsmath,amstext,amsfonts,amsbsy,amssymb,amscd,bbm,epsfig}

\def\be{\begin{equation}}
\def\ee{\end{equation}}
\def\bc{\begin{center}}
\def\ec{\end{center}}
\def\bea{\begin{eqnarray}}
\def\eea{\end{eqnarray}}
\newcommand{\bi}{\begin{itemize}}
\newcommand{\ei}{\end{itemize}}                 

\newcommand{\ba}{\begin{array}{c}}
\newcommand{\bad}{\begin{array}{ccc}}
\newcommand{\ea}{\end{array}}

\newcommand{\ov}{\overline}

\newcommand{\bra}[1]{\ensuremath{\langle #1 |}}   
\newcommand{\ket}[1]{\ensuremath{| #1 \rangle}}   
\newcommand{\eps}{\varepsilon}
\newcommand{\ldm}{\ensuremath{{\Delta m_{31}^2}}}          

\allowdisplaybreaks[4] 

\begin{document}

\begin{titlepage}

\renewcommand{\thefootnote}{\alph{footnote}}

\vspace*{-3.cm}
\begin{flushright}
RM3-TH/14-3\\
SISSA 14/2014/FISI
\end{flushright}
\vspace*{0.7cm}

\vspace*{0.2cm}

\title{Constraining new physics scenarios in neutrino oscillations from Daya Bay data}

\author{I.~Girardi}
\affiliation{SISSA/INFN, Via Bonomea 265, I-34136 Trieste, Italy }
\author{D.~Meloni}
\affiliation{Dipartimento di Matematica e Fisica, 
Universit\`{a} di Roma Tre, Via della Vasca Navale 84, I-00146 Rome}

\date{\today}

\begin{abstract}

We perform for the first time a detailed fit to the $\bar \nu_e \to \bar \nu_e$ disappearance data of the Daya Bay 
experiment to constrain the parameter 
space of models where sterile neutrinos can propagate in a 
large compactified extra dimension (LED) and models where non-standard interactions affect the neutrino production and 
detection (NSI). We find that 
the compactification radius $R$ in LED scenarios can be constrained at the level of $0.57 \, \mu m$ for normal ordering and
of $0.19\, \mu m$ for inverted ordering, at 2$\sigma$ confidence level.
For the NSI model, reactor data put a strong upper bound on the parameter 
$\eps_{ee}$ at the level of $\sim 10^{-3}$, whereas the main effect of $\eps_{e\mu}$ and $\eps_{e\tau}$
is a worsening of the determination of $\theta_{13}$.
\end{abstract}

\maketitle

\end{titlepage}

After the recent measure of the reactor angle by T2K \cite{Abe:2013hdq}, Daya Bay \cite{An:2013uza} and Reno \cite{Ahn:2012nd} experiments, 
the standard picture of neutrino oscillation seems now to be very well established, with only few items 
to be clarified, namely the presence of CP violation in the PMNS mixing matrix and 
the ordering of the mass eigenvalues. Beyond this standard picture, the possibility that new physics can affect 
neutrino oscillation is not excluded and, although expected to be small, deserve a closer look. 
A popular interesting model of new physics in neutrino oscillations is the one where sterile neutrinos can propagate, 
as well as gravity, in 
large $\delta$ compactified extra dimensions (LED) \cite{ADD} whereas the Standard Model (SM) left-handed
neutrinos are confined to a four-dimensional spacetime brane \cite{Barbieri,Mohapatra,Davoudiasl:2002fq}.
Experiments based on the torsion pendulum instrument
set an upper limit on the largest compactification radius
$R < 37 \mu m$ for $\delta = 2$ at $95$\% CL
\cite{Beringer:1900zz}.
Much stronger bounds can be set by astrophysics \cite{Hannestad:2003yd}
but they are not completely model independent, so an analysis of the 
constraints coming from neutrino oscillation data still deserves a lot of attention.
Since scenarios with only one extra dimensions have been already ruled-out \cite{Beringer:1900zz},
we assume to work with an effective 5-dimensional theory in which only the radius $R$ of the largest new dimension 
is the relevant parameter for neutrino oscillation. 
Under these assumptions, the transition amplitude $\overline \nu_e \rightarrow \overline \nu_e$ in vacuum is given by \cite{Davoudiasl:2002fq}:
\begin{eqnarray}
A_{e e} (L) = \sum_{i = 1}^3 \sum_{n = 0}^{\infty}U^{e i} U^{* e i} \left[ U_i^{0n} \right]^2 
\exp \left( i \frac{\lambda_i^{(n)2} L}{2 E_{\nu} R^2}\right), \,\,
\end{eqnarray}
where $U^{e i}$ is the first row of the $U_{PMNS}$ matrix, 
$\lambda_i^{(n)}$ are the eigenvalues of the neutrino mass matrix given by 
$\lambda^{(n)}_i \simeq \xi_i / \sqrt{2}$ for $n = 0$  and $\lambda_i \simeq n + \xi^2_i / (2 n)$
 for $n \geq 1$, with $\xi_i \equiv \sqrt{2} m_i R$ ($m_i = $ absolute neutrino masses) and
$U_i^{0n}$ are the elements of the matrix
describing the transition between the zero mode and the n-th Kaluza-Klein states \cite{Davoudiasl:2002fq},
$\left( U_i^{0n} \right)^2 \simeq \xi^2_i / n^2$.
For the normal 
ordering (NO) we assume $m_3 > m_2 > m_1=m_0$,
whereas for the inverted ordering (IO) $m_2> m_1 > m_3=m_0$.
We note that the effect of LED is significant more pronounced in IO than NO
because the latter amplitude $A_{ee}^{NO}(L) \sim \xi_1^2 \,U_{e1}^2 + \xi_2^2 \,U_{e2}^2+ \xi_3^2 \,U_{e3}^2$
is dominated by $\xi_3^2 \,\sin^2 \theta_{13}$, then suppressed by $\theta_{13}$, whereas in the 
IO $A_{ee}^{IO}(L) \sim \xi_1^2 \,U_{e1}^2 + \xi_2^2 \,U_{e2}^2$ and does not suffer of such a suppression.
We then expect the IO scenario to give better constraints on $R$ and $m_0$ than the NO case.
\\
Another interesting model of physics beyond the standard three neutrino oscillation is the one called non-standard neutrino interactions (NSI) \cite{Wolf78}, 
in which new physics effects can appear at low energy in terms of unknown couplings $\varepsilon_{\alpha\beta}$, generated after integrating out 
new degrees of freedom, with very large mass scales. 
In reactor experiments, the new couplings can affect neutrino 
production ("s") and detection ("d") \cite{Grossman:1995wx}, so
the neutrino states are a superposition of pure orthonormal flavor eigenstates
\cite{Ohlsson:2013nna,Meloni:2009cg} according to: $|\nu^{\rm s}_e \rangle  = \big[ (1 + \varepsilon^{ s}) |\nu \rangle \big]_{e}$
and $\langle \nu^{\rm d}_e|  =  \big[ \langle \nu
|  (1 + \varepsilon^{ d}) \big]_{e}$,
with $\varepsilon^{s}$ and $\varepsilon^{d}$ generic non-unitary 
transformations.
Since the parameters $\eps_{e\alpha}^s$ and  $\eps_{\alpha e}^{d}$ receive contributions from 
the same higher dimensional operators \cite{Kopp:2007ne}, one can constrain them by the relation
$
\eps_{e\alpha}^s = \eps_{\alpha e}^{d*} 
\equiv  \eps_{e\alpha} e^{{\rm i}  \, \phi_{e\alpha}}\;,
$
being $\eps_{e\alpha}$ the modulus and $\phi_{e\alpha}$ the argument of $\eps_{e\alpha}^s$.
The oscillation probability $P_{ee} \equiv P(\overline \nu_e \rightarrow \overline \nu_e)$ up to $O(\eps)$
can be obtained by squaring the amplitude $\bra{\nu^d_e} e^{-i H L} \ket{\nu^s_e}$:
\begin{eqnarray}
&& P_{ee}  = 1 
 - \sin^2 2\theta_{13} \sin^2 \Delta + 4 \eps_{ee} \cos \phi_{ee} \label{Eq:PNSIlimit} \\
&& -4 \eps_{e\mu}  \sin 2\theta_{13} \sin \theta_{23} \cos 2 \theta_{13}   \cos(\delta - \phi_{e\mu}) \sin^2 \Delta  \nonumber \\
&&  -4 \eps_{e\tau} \sin 2\theta_{13} \cos \theta_{23}  \cos 2 \theta_{13}  \cos(\delta - \phi_{e\tau} ) \sin^2 \Delta \nonumber \,, 
\end{eqnarray}
where $\Delta \equiv  \left[ \frac{\ldm \, L}{4 E_{\nu}} \right ]$, with $L$ being the source-to-detector distance,
$E_{\nu}$ the neutrino energy and $\ldm = m_3^2 - m_1^2$.
In the first line of Eq.~(\ref{Eq:PNSIlimit}) we can recognize the "zero-distance" term driven by $\eps_{ee}$, 
which gives a non vanishing contribution even in the limit of very small $L/E_\nu$. In addition,  $\eps_{e\mu}$ 
and  $\eps_{e\tau}$ appear with only slightly different coefficients.
However, contrary to what happens for $\eps_{ee}$, $\eps_{e\mu,\tau}$ exhibit a strong correlation with the 
reactor angle which, on the one hand, does not allow to set any stringent bound on them and, on the other hand,
can worsen the extraction of $\theta_{13}$ and $\ldm$ from the data~\cite{Ohlsson:2008gx,Leitner:2011aa}.
A model-independent analysis \cite{enrique} has shown that all bounds 
on production and detection NSI's are at the level of $10^{-2}$: 
$\varepsilon_{ee} < 0.041$,  $\varepsilon_{e\mu} < 0.025$
and $\varepsilon_{e\tau} < 0.041$,
whereas for the CP violating phases no constraints are known.

In this paper we make use of the recent $\bar \nu_e \to \bar \nu_e$ disappearance data of the Daya Bay experiment 
to constrain the parameter space of NSI and LED scenarios. 
Our main results are that neutrino oscillation data can provide strong upper bounds on $\varepsilon_{ee}$
at the level of ${\cal O}(10^{-3})$, 
whereas for $R$ the exclusion limits are
between 1 and 2 order of magnitudes below the limits quoted in \cite{Beringer:1900zz}.

The Daya Bay experimental setup we take into account \cite{An:2013uza} consists of six antineutrino detectors (ADs)
and six reactors, D1, D2, L1, L2, L3, L4.  
The antineutrino spectra emitted by the nuclear reactors have been recently 
estimated in \cite{Mueller:2011nm,Huber:2011wv}. For each AD's, the flux of arriving 
$\ov \nu_e$ has contributions from the isotopes $ $$^{235}$U, $^{239}$Pu, $^{238}$U, and $^{241}$Pu, 
with weights reported in \cite{talk:DayaBay};
for a given isotope, we adopt the convenient parametrization of \cite{Mueller:2011nm}.
For our analysis we used the data set accumulated during 217 days 
reported in \cite{An:2013zwz}, where the detected antineutrino candidates are collected in the far hall, EH3, and 
in the near halls EH1, EH2. A bin-to bin normalization 
has been fixed in order to reproduce the unoscillated rates.
The antineutrino energy $E_{\, \ov \nu_e}$ is reconstructed
by the prompt energy deposited by the positron $E_{ \rm prompt}$ using 
the approximated relation \cite{An:2013uza}:
$E_{\, \ov \nu_e} \simeq E_{\rm prompt} + 0.8 \; {\rm MeV}$.
The energy resolution
function is a Gaussian function 
with  $\sigma(E)  [{\rm MeV}] = 0.08  \sqrt{E_{\nu}/{\rm MeV} - 0.8}$.
The antineutrino cross section for the inverse beta decay (IBD) process has been taken
from \cite{Vogel:1999zy}.
In order to perform a
proper statistical treatment of correlations and degeneracy, we used a modified version of the GLoBES software
\cite{GLOB2} and construct an adequate definition of the $\chi^2$ function \cite{An:2013uza}:
\begin{eqnarray}
 \label{eqn:chispec}
&& \chi^2(\theta,\Delta m^2, \vec S,\alpha_r,\varepsilon_d,\eta_d)  = \nonumber \\
&& \sum_{d=1}^{6}\sum_{i=1}^{36}
 \frac{\left[M_i^d -T_i^d \cdot \left(1 +
 \sum_r\omega_r^d\alpha_r + \varepsilon_d  \right) 
  + \eta_d \right]^2}
 {M_i^d + B_i^d } \nonumber \\
&& + \sum_r\frac{\alpha_r^2}{\sigma_r^2}
 + \sum_{d=1}^{6}\left[
\frac{\varepsilon^2_d}{\sigma^2_d}
 + \frac{\eta_d^2}{\sigma_{B_d}^2}
 \right]   + \mbox{Priors} \,. 
\end{eqnarray}
In the previous formula, $\vec S$ is a vector containing the new physics parameters,
$M^d_i$ are the measured IBD events of the d-th detector ADs
in the i-th bin, $B^d_i$ the corresponding background and $T^d_i = T_i(\theta,\Delta m^2, \vec S)$ are the
theoretical prediction for the rates. The parameter $\omega_r^d$ is the fraction
of IBD contribution of the r-th reactor to the d-th detector AD, determined by
the approximated relation $\omega_r^d \sim L_{rd}^{-2} / (\sum_{r = 1}^6 1/L_{rd}^2 )$,
where $L_{rd}$ is the distance between the d-th detector and the r-th reactor. The parameter $\sigma_d$
is the uncorrelated detection uncertainty ($\sigma_d = 0.2$\%) and $\sigma_{B_d}$ is the background uncertainty of the d-th detector 
obtained using the information given in \cite{An:2013zwz}:
$\sigma_{B_1} = \sigma_{B_2} = 8.21$, $\sigma_{B_3} = 5.95$, $\sigma_{B_4} = \sigma_{B_5} = \sigma_{B_6} = 1.15$.
Eventually, $\sigma_r = 0.8$\% is the correlated reactor uncertainties. 
The corresponding pull parameters are ($\varepsilon_d,\eta_d,\alpha_r$).
The main relevant point in this discussion is which priors must be implemented in the fitting function. 
Since Daya Bay has measured $\theta_{13}$ with very high precision, we cannot use its determination to constraint the reactor angle
when fitting the new physics parameters, otherwise we would use the same data twice. Similar 
considerations can also be done for the atmospheric mass difference, which primarily drives 
the standard oscillation term in  $P_{ee}$.
So, when studying LED in the plane ($R,m_0$) and NSI in the plane $(\eps_{\alpha\beta},\phi_{\alpha\beta})$,
we adopt the following strategy: we do not impose any constraints of $\theta_{13}$ and
we set the uncertainty on $\ldm$ at values larger than the current determination:
$
\Delta m_{31}^2 = (2.35 \pm 10 \%) \times
10^{-3} \; {\rm eV}^2  
$ 
(we carefully checked that leaving $\ldm$ completely unconstrained our results do not change).
For the atmospheric angle and the solar parameters the situation is a bit different since the standard probability does not depend on them;
however, they couple to the new physics parameters, both in LED and NSI scenarios, so
we need to impose  external constraints, chosen as follows \cite{Capozzi:2013csa}: $\sin^2 \theta_{23}=0.425 \pm 0.029$ for NO and $\sin^2 \theta_{23}=0.437 \pm 0.173$ for IO, $\Delta m_{21}^2 =(7.54 \pm 0.26)\times 10^{-5} \; {\rm eV}^2$
and $\sin^2 \theta_{12} = 0.308 \pm 0.017$.
Whenever necessary, the standard CP violating phase $\delta$ will be considered as a free parameter.
\\
The results in the standard $[\sin^2 2 \theta_{13},\ldm]$-plane, instead,
are obtained marginalizing also over  $R$ and $m_0$ for LED and over
$\eps$ and $\phi$ for NSI,   in the perturbative regions identified by
$\xi_i \equiv \sqrt{2} m_i R < 0.2$ and $\eps < 0.041$, while $\phi \in [0,2\pi]$.

\begin{figure*}[t!]
  \includegraphics[height=6cm]{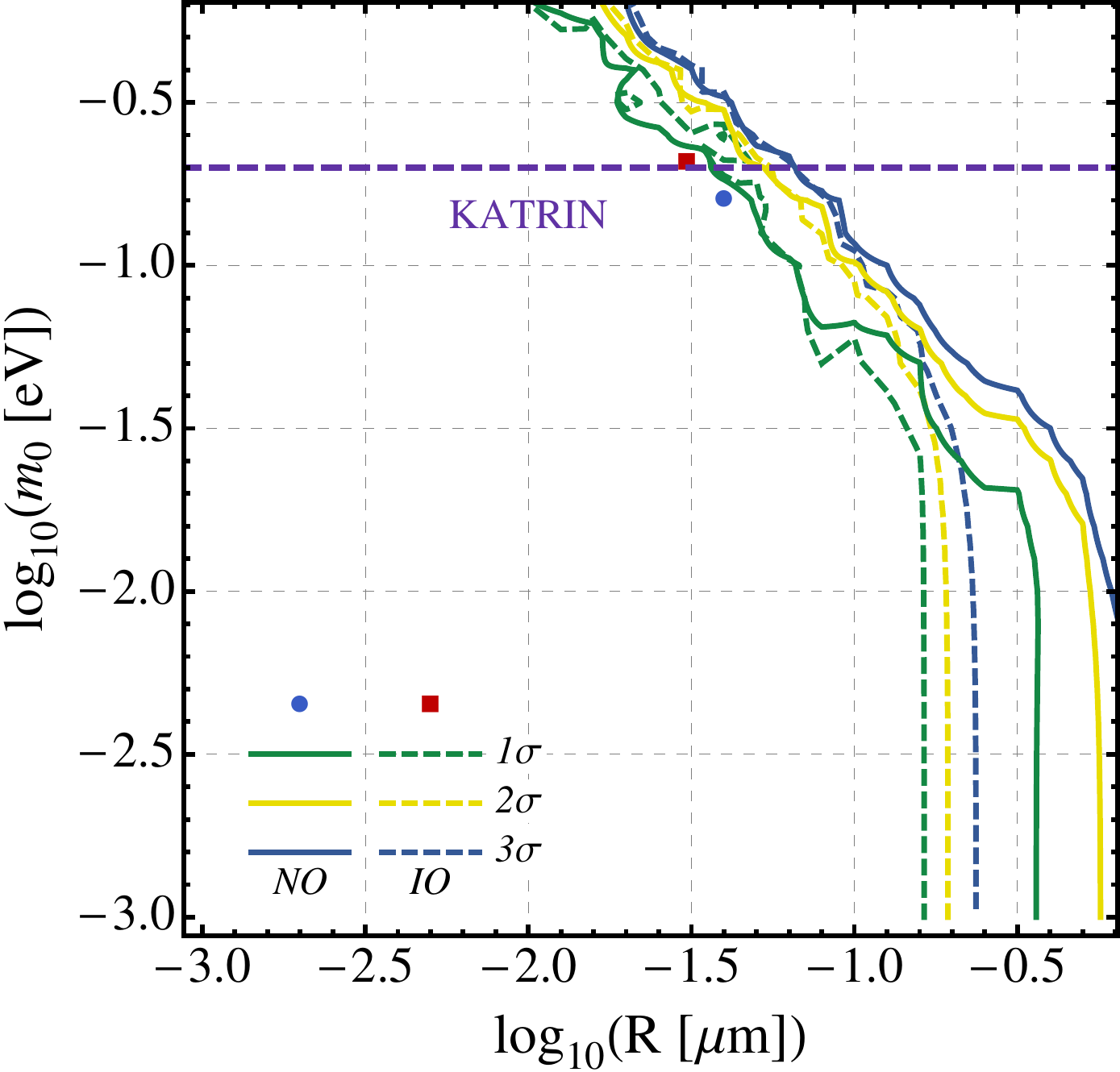}\hspace{2ex}%
  \includegraphics[height=6cm]{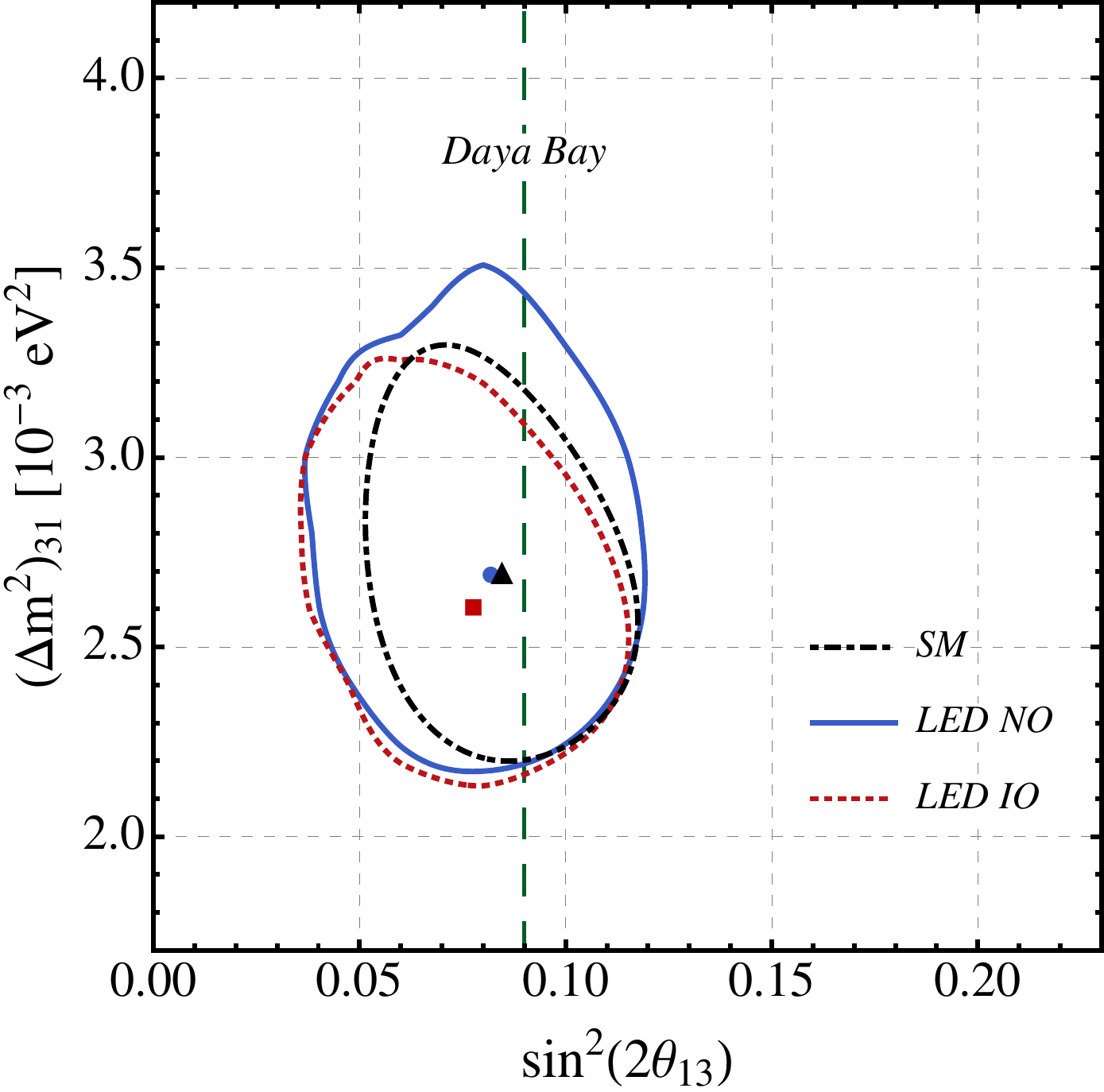}
  \caption{ \it Left panel: allowed regions for NO and IO LED model in the $[\log_{10}( R ),\log_{10}( m_0 )]$-plane
at 1$\sigma$, 2$\sigma$ and 3$\sigma$ CL. The best fit points for both hierarchies are indicated with a 
circle (NO) and a square (IO). Right panel: 3$\sigma$ CL in the $[\sin^2 (2 \theta_{13}), \ldm]$-plane.
The best fit points are indicated with a circle (LED NO), a square (LED IO) and a triangle (SM). 
The dashed vertical line represents the value of the $\theta_{13}$ quoted 
by the Daya Bay collaboration \cite{An:2013zwz}.
}
\label{fig:m0RH}
\end{figure*}

We first consider the bounds on the size of the large extra dimension $R$ and on the lightest neutrino mass, in the $[R, m_0]$-plane.
Our results are shown in left panel of Fig.~\ref{fig:m0RH}, where we displayed the 1, 2 and 3$\sigma$ CL regions.
Both ordering of the neutrino masses, and the related values of the $\chi^2_{min}/$dof, have been considered; solid lines refer to
the NO whereas the dashed ones refer to the IO. The horizontal dashed line represents the future upper limit on $m_0$ from the $\beta$-decay experiment KATRIN \cite{Eitel:2005hg}. 
%
Since the standard oscillation physics already gives a good fit to the data, small
values of  $R$ and $m_0$ are obviously allowed; the correlation existing among these parameters, however,
is quite strong and excludes large values of $R$ and $m_0$.
In particular, bounds on the compactification radius can be set at the level 
of some units of $10^{-1}$ $\mu m$:
$R <  0.36\; (0.16)$ at 1$\sigma$, $R < 0.57 \; (0.19)$ at 2$\sigma$ and  $R < $ None  (0.23) at 3$\sigma$
for NO (IO).
The best fit points (a circle for NO and a square for IO in Fig.~\ref{fig:m0RH}) and the related $\chi^2_{min}/\mathrm{dof}$ have the following values:
$R [\mu m] = 0.04\, (0.032)$, $m_0 [{\rm eV}]   = 0.16 \,(0.20)$ with $\chi^2_{min}/\mathrm{dof}  = 45/106\, (45/106)$,
where the numbers in parenthesis refer to the IO.
However, it is worth to mention that they only have an indicative 
meaning, since the $\chi^2$ function is almost flat in the allowed regions.
\\
It is an interesting question to check whether the new physics parameters introduce some bias in the simultaneous extraction of $\theta_{13}$ and $\ldm$.
In the right panel of Fig.~\ref{fig:m0RH} we show the 3$\sigma$
CL allowed region in the $[\sin^2 (2 \theta_{13}), \ldm]$-plane for NO (solid line), IO (dotted line),
and the standard model (dot-dashed line) results. We can appreciate an increase of the allowed $\theta_{13}$ and 
$\Delta m^2_{31}$ 3$\sigma$ CL regions, at the level of 25\% toward smaller reactor angles and 5\% to larger masses.
In Tab.~\ref{tab:LED_fitDB} we summarise the obtained results, reporting the best fit values and 1$\sigma$ errors
for $\sin^2 2 \theta_{13}$, $\Delta m^2_{31}$ and the related value $\chi^2_{min}/$dof for the three scenarios shown in Fig.~\ref{fig:m0RH}.
\begin{table}[h]
\begin{center}
\begin{tabular}{|c | c  | c |  c |}
\hline
\bf  Parameter & \bf SM & \bf LED NO & \bf LED IO  \\
\hline
 $\sin^2 2 \theta_{13}$ &  $0.085^{+0.015}_{-0.016}$  & $0.082^{+0.021}_{-0.022}$ & $0.078^{+0.018}_{-0.018}$  \\
  $\Delta m^2_{31}/10^{-3}\ [{\rm eV}^2]$ & $2.69^{+0.27}_{-0.24}$  & $2.69^{+0.30}_{-0.25}$ & $2.60^{+0.24}_{-0.20}$ \\
 $\chi^2_{min}/\mathrm{dof}$ &  $43/106$  & $43/106$ & $42/106$ \\
\hline
\end{tabular}
\caption{\it Best fit points and 1$\sigma$ errors for
$\sin^2 2 \theta_{13}$, $\Delta m^2_{31}$ and the value of $\chi^2_{min}/$dof.
Results are for the SM, the LED NO and LED IO cases.}
\label{tab:LED_fitDB}
\end{center}
\end{table}

\begin{figure*}[t!]
  \includegraphics[height=6cm]{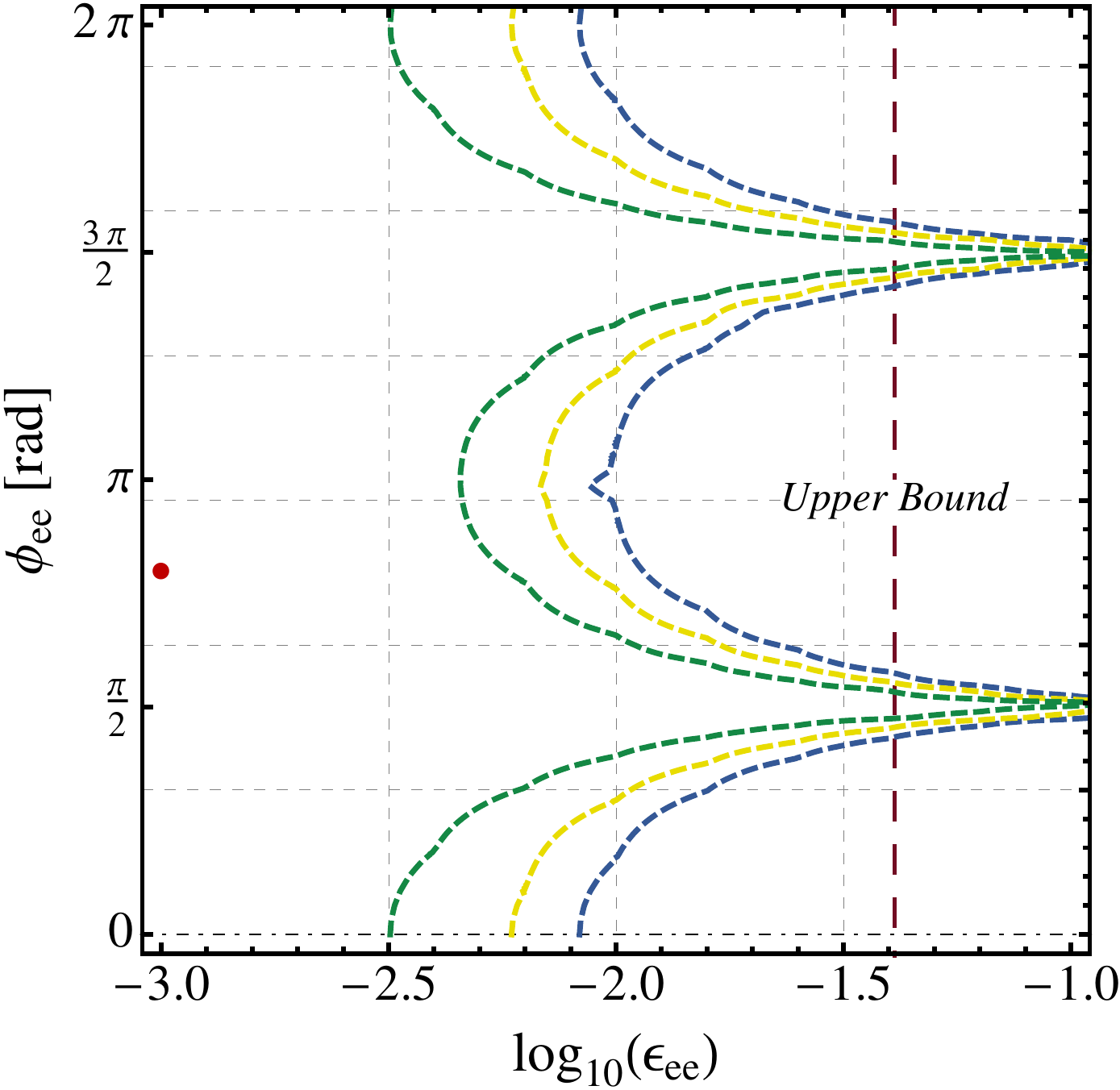}\hspace{2ex}%
  \includegraphics[height=6cm]{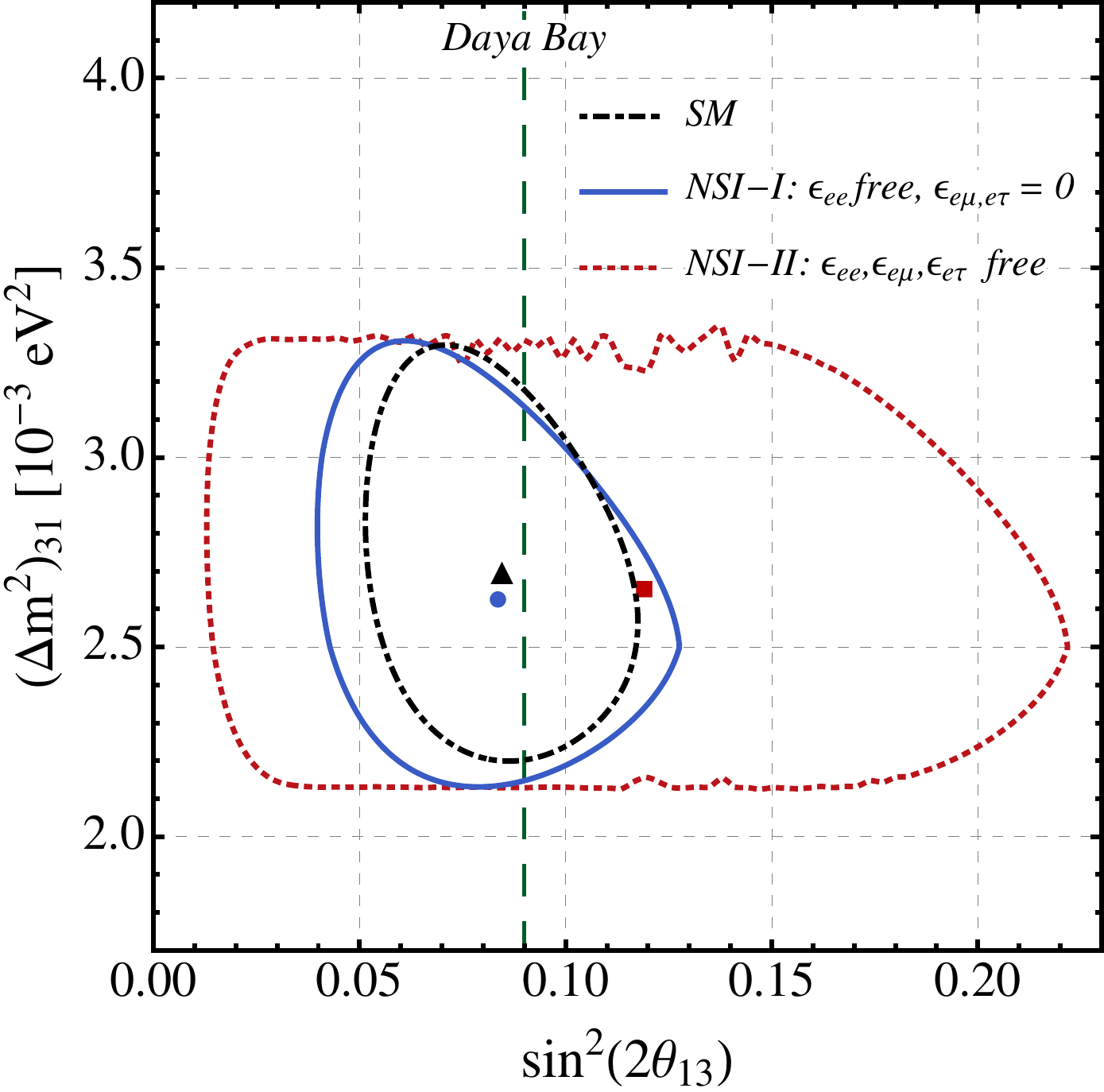}
  \caption{ \it Left panel: excluded regions in the $[\eps_{ee},\phi_{ee}]$-plane at 1, 2 and 3$\sigma$ CL. 
The vertical line corresponds to $\eps = 0.041$. Circles are the obtained best fit points.
Right Panel: 3$\sigma$ CL in the $[\sin^2 (2 \theta_{13}),\ldm]$-plane for the SM (dot-dashed),
for $\eps_{e\mu} = \eps_{e\tau} = 0$ (NSI-I solid) and for free parameters (NSI-II dotted).
The best fit points are indicated with a circle (NSI-I), a square (NSI-II) and a triangle (SM).}
\label{fig:epsphi}
\end{figure*}

The parameter space for the NSI investigation is larger than for LED,
 consisting of the moduli $\eps_{ee}$,
$\eps_{e\mu}$, $\eps_{e\tau}$ and the new CP phases $\phi_{ee}$,
$\phi_{e\mu}$, $\phi_{e\tau}$. 
The  study of the allowed regions in the $[\eps_{ee},\phi_{ee}]$-plane is performed marginalizing over all the parameters,
including $\eps_{e\mu}$, $\eps_{e\tau}$ and their phases.
The result of such a procedure is presented 
in the left panel of Fig.~\ref{fig:epsphi}, where the 1, 2 and 3$\sigma$ CL have been displayed, together with the 
obtained best fit point (circle). 
The vertical dashed line is at $\eps_{ee}=0.041$.
%
The results can be easily explained using the approximate probability in Eq.~(\ref{Eq:PNSIlimit}):
a maximal sensitivity to $\eps_{ee}$ is obtained in correspondence of vanishing CP phase;
in addition, the symmetry 
around $\phi_{ee} \sim \pi$ is a trivial consequence of the $\cos \phi_{ee}$ dependence in $P_{ee}$.
Upper bounds on $\eps_{ee}$ can be set of the order $\eps_{ee} \lesssim 3 \cdot 10^{-3}$ for $\phi_{ee} \sim 0, 2\pi$, 
and $\eps_{ee} \lesssim 5 \cdot 10^{-3}$ for $\phi_{ee} \sim \pi$; in both cases, the Daya Bay data significantly 
lower the existing upper limit
on $\eps_{ee}$. The best fit point is:
$\eps_{ee} =   0.001$, $\phi_{ee} = 2.5$, with $\chi^2_{min}/\mathrm{dof} = 45/106$.
We have checked that, contrary to what
described above, the correlation among $\theta_{13}$ and $\eps_{e\mu}$, $\eps_{e\tau}$
does not allow to set any interesting bounds on them.
\\
As for the LED case, we are interested in estimating the determination of the reactor angle $\theta_{13}$
and the mass difference $\ldm$ in presence of 
NSI. Driven by the considerations after Eq.~(\ref{Eq:PNSIlimit}), 
we study two different cases, both illustrated in the right panel of Fig.~\ref{fig:epsphi}; in the case NSI-I (solid line), we set $\eps_{e\mu}=\eps_{e\tau}=0$ 
and marginalize over  $\eps_{ee}$ and $\phi_{ee}$. We see  that 
the effect is a variation in the determination of $\sin^2 2 \theta_{13}$ of some $\sim 30$\%, and 
of the mass difference around $\sim 5$\%. 
\\
In the case NSI-II (dotted line), we also leave $\eps_{e\mu}$ and $\eps_{e\tau}$ and the related CP phases as free parameters; 
the impact on the determination of $\theta_{13}$ is really large: beside a drift of the best fit point toward larger values, 
the allowed 3$\sigma$ interval covers a broader 3$\sigma$ range, $0.013\lesssim \sin^2 (2 \theta_{13})\lesssim 0.22$ at 3$\sigma$.
The obtained best fit points and the 1$\sigma$ errors 
for NSI-I and NSI-II are summarised in Tab.~\ref{tab:NSI_fitDB}.
\begin{table}[h]
\begin{center}
\begin{tabular}{|c | c  | c |  c |}
\hline
\bf  Parameter & \bf SM & \bf NSI-I & \bf NSI-II  \\
\hline
 $\sin^2 2 \theta_{13}$ &  $0.085^{+0.015}_{-0.016}$  & $0.084^{+0.022}_{-0.021}$ & $0.119^{+0.08}_{-0.09}$  \\
  $\Delta m^2_{31}/10^{-3}\ [{\rm eV}^2]$ & $2.69^{+0.27}_{-0.24}$  & $2.62^{+0.30}_{-0.22}$ & $2.65^{+0.27}_{-0.25}$ \\
 $\chi^2_{min}/\mathrm{dof}$ &  $43/106$  & $43/106$ & $43/106$ \\
\hline
\end{tabular}
\caption{\it Best fit points and 1$\sigma$ errors for
$\sin^2 2 \theta_{13}$, $\Delta m^2_{31}$ and the value of $\chi^2_{min}/$dof.
Results are for the SM, the NSI-I and NSI-II cases.}
\label{tab:NSI_fitDB}
\end{center}
\end{table}

In summary, the most recent data of the Daya Bay experiment \cite{An:2013zwz}
allow to set strong upper bounds on the new physics parameters involved in LED and NSI scenarios.
For the compactification radius $R$, the limits at 2$\sigma$ are
$R <  0.19$ $\mu$m for IO and $R < 0.57$ $\mu$m for NO, much stringent that the current limits from torsion pendulum
experiments. For the NSI case, a special role is played by the $\eps_{ee}$ parameter since it is not correlated to
$\theta_{13}$. The experimental data set a strong upper bound
of ${\cal O}(10^{-3})$ at 3$\sigma$. On the other hand, $\eps_{e\mu}$ and $\eps_{e\tau}$ suffer from a strong
correlation to $\theta_{13}$ and, therefore, no significant
sensitivity has been found. However, they play a major role in the determination of $\theta_{13}$ and $\ldm$;
our analysis shows that, even assuming $\eps_{ee}=0$,  the allowed regions for $\theta_{13}$ are 
much larger than the SM ones; in addition,  the best fit value for $\theta_{13}$ is driven to values larger by roughly
40\%.
On the other hand, the determination of the squared
mass difference $\ldm$ is less affected by this type of new physics and the fit procedures return
values  very similar to the SM case.

We acknowledge MIUR (Italy) for
financial support under the program Futuro in Ricerca 2010 (RBFR10O36O). We are strongly indebted with Patrick Huber
and Camillo Mariani for useful discussions on the subtleties of the Daya Bay and Double Chooz experiments and
S.T. Petcov for useful suggestions.
 This work was supported in part by the INFN program on
 ``Astroparticle Physics'' and  by the European Union FP7-ITN INVISIBLES
 (Marie Curie Action PITAN-GA-2011-289442-INVISIBLES) (I.G.).

\expandafter\ifx\csname natexlab\endcsname\relax\def\natexlab#1{#1}\fi
\expandafter\ifx\csname bibnamefont\endcsname\relax
  \def\bibnamefont#1{#1}\fi
\expandafter\ifx\csname bibfnamefont\endcsname\relax
  \def\bibfnamefont#1{#1}\fi
\expandafter\ifx\csname citenamefont\endcsname\relax
  \def\citenamefont#1{#1}\fi
\expandafter\ifx\csname url\endcsname\relax
  \def\url#1{\texttt{#1}}\fi
\expandafter\ifx\csname urlprefix\endcsname\relax\def\urlprefix{URL }\fi
\providecommand{\bibinfo}[2]{#2}
\providecommand{\eprint}[2][]{\url{#2}}


\begin{thebibliography}{27}


\bibitem[{\citenamefont{K.~Abe}(2014)}]{Abe:2013hdq}
\bibinfo{author}{\bibfnamefont{K.}~\bibnamefont{Abe}} \bibnamefont{et~al.}
  (\bibinfo{collaboration}{T2K Collaboration}), \bibinfo{journal}{Phys.
  Rev. Lett.} \textbf{\bibinfo{volume}{112}}, \bibinfo{pages}{061802}
  (\bibinfo{year}{2014}). 
  
  
 \bibitem[{\citenamefont{F.~P.~An}(2013)}]{An:2013uza}
\bibinfo{author}{\bibfnamefont{F.}~\bibfnamefont{P.}~\bibnamefont{An}} \bibnamefont{et~al.}
  (\bibinfo{collaboration}{Daya Bay Collaboration}), \bibinfo{journal}{Chin.
  Phys. C.} \textbf{\bibinfo{volume}{37}}, \bibinfo{pages}{011001}
  (\bibinfo{year}{2013}). 


 \bibitem[{\citenamefont{J.~K.~Ahn}(2012)}]{Ahn:2012nd}
\bibinfo{author}{\bibfnamefont{J.}~\bibfnamefont{K.}~\bibnamefont{An}} \bibnamefont{et~al.}
  (\bibinfo{collaboration}{RENO Collaboration}), \bibinfo{journal}{Phys.
  Rev. Lett.} \textbf{\bibinfo{volume}{108}}, \bibinfo{pages}{191802}
  (\bibinfo{year}{2012}). 


  \bibitem[{\citenamefont{N.~Arkani-Hamed}(1998)}]{ADD}
\bibinfo{author}{\bibfnamefont{N.}~\bibfnamefont{Arkani-Hamed}, \bibfnamefont{S.}~\bibnamefont{Dimopoulos}
 and \bibfnamefont{G.}~\bibnamefont{Dvali}, 
  \bibinfo{journal}{Phys. Lett. B} \textbf{\bibinfo{volume}{429}}, \bibinfo{pages}{263}}
  (\bibinfo{year}{1998}); 
  \bibinfo{author}{\bibfnamefont{I.}~\bibnamefont{Antoniadis}, \bibfnamefont{N.}~\bibfnamefont{Arkani-Hamed}, \bibfnamefont{S.}~\bibnamefont{Dimopoulos}
 and \bibfnamefont{G.}~\bibnamefont{Dvali}, 
     \bibinfo{journal}{Phys. Lett. B} \textbf{\bibinfo{volume}{436}}, \bibinfo{pages}{257}}
  (\bibinfo{year}{1998});
 \bibinfo{author}{\bibfnamefont{N.}~\bibfnamefont{Arkani-Hamed}, \bibfnamefont{S.}~\bibnamefont{Dimopoulos}
 and \bibfnamefont{G.}~\bibnamefont{Dvali}, 
     \bibinfo{journal}{Phys. Rev. D} \textbf{\bibinfo{volume}{59}}, \bibinfo{pages}{086004}}
  (\bibinfo{year}{1999}).


 \bibitem[{\citenamefont{R.~Barbieri}(2000)}]{Barbieri}
\bibinfo{author}{\bibfnamefont{R.}~\bibnamefont{Barbieri}, \bibfnamefont{P.}~\bibnamefont{Creminelli}
 and \bibfnamefont{A.}~\bibnamefont{Strumia}, 
  \bibinfo{journal}{Nucl. Phys. B} \textbf{\bibinfo{volume}{585}}, \bibinfo{pages}{28}}
  (\bibinfo{year}{2000}). 


   \bibitem[{\citenamefont{R.~N.~Mohapatra}(1999)}]{Mohapatra}
\bibinfo{author}{\bibfnamefont{R.}~\bibfnamefont{N.}~\bibfnamefont{Mohapatra}, \bibfnamefont{S.}~\bibnamefont{Nandi}
 and \bibfnamefont{A.}~\bibfnamefont{Perez-Lorenzana}, 
  \bibinfo{journal}{Phys. Lett. B} \textbf{\bibinfo{volume}{466}}, \bibinfo{pages}{115}}
  (\bibinfo{year}{1999}); 
  \bibinfo{author}{\bibfnamefont{R.}~\bibfnamefont{N.}~\bibfnamefont{Mohapatra}
 and \bibfnamefont{A.}~\bibfnamefont{Perez-Lorenzana}, 
  \bibinfo{journal}{Nucl. Phys. B} \textbf{\bibinfo{volume}{576}}, \bibinfo{pages}{466}}
  (\bibinfo{year}{2000}); 
    \bibinfo{author}{\bibfnamefont{R.}~\bibfnamefont{N.}~\bibfnamefont{Mohapatra}
 and \bibfnamefont{A.}~\bibfnamefont{Perez-Lorenzana}, 
  \bibinfo{journal}{Nucl. Phys. B} \textbf{\bibinfo{volume}{593}}, \bibinfo{pages}{451}}
  (\bibinfo{year}{2001}). 

  
   \bibitem[{\citenamefont{H.~Davoudiasl}(2002)}]{Davoudiasl:2002fq}
\bibinfo{author}{\bibfnamefont{H.}~\bibnamefont{Davoudiasl}, \bibfnamefont{P.}~\bibnamefont{Langacker}
 and \bibfnamefont{M.}~\bibnamefont{Perelstein}, 
  \bibinfo{journal}{Phys. Rev. D} \textbf{\bibinfo{volume}{65}}, \bibinfo{pages}{105015}}
  (\bibinfo{year}{2002}). 
  
  
  \bibitem[{\citenamefont{J.~Beringer}(2012)}]{Beringer:1900zz}
\bibinfo{author}{\bibfnamefont{J.}~\bibnamefont{Beringer}} \bibnamefont{et~al.}
  (\bibinfo{collaboration}{Particle Data Group Collaboration}), \bibinfo{journal}{Phys.
  Rev. D} \textbf{\bibinfo{volume}{86}}, \bibinfo{pages}{010001}
  (\bibinfo{year}{2012}). 
  

   \bibitem[{\citenamefont{S.~Hannestad}(2003)}]{Hannestad:2003yd}
\bibinfo{author}{\bibfnamefont{S.}~\bibnamefont{Hannestad}
 and \bibfnamefont{G.}~\bibfnamefont{G.}~\bibnamefont{Raffelt}, 
  \bibinfo{journal}{Phys. Rev. D} \textbf{\bibinfo{volume}{67}}, \bibinfo{pages}{125008}}
  (\bibinfo{year}{2003});
  \bibinfo{author}{Erratum-ibid.
   \bibinfo{journal}{Phys. Rev. D} \textbf{\bibinfo{volume}{69}}, \bibinfo{pages}{029901}}
  (\bibinfo{year}{2004}). 

  
  \bibitem[{\citenamefont{L.~Wolfenstein}(1978)}]{Wolf78}
\bibinfo{author}{\bibfnamefont{L.}~\bibnamefont{Wolfenstein},
  \bibinfo{journal}{Phys. Rev. D} \textbf{\bibinfo{volume}{17}}, \bibinfo{pages}{2369}}
  (\bibinfo{year}{1978});
\bibinfo{author}{\bibfnamefont{M.}~\bibfnamefont{M.}~\bibnamefont{Guzzo}, \bibfnamefont{A.}~\bibnamefont{Masiero} 
and \bibfnamefont{S.}~\bibfnamefont{T.}~\bibnamefont{Petcov}},
 \bibinfo{journal}{Phys. Lett. B} \textbf{\bibinfo{volume}{260}}, \bibinfo{pages}{154}
  (\bibinfo{year}{1991}).
 

 \bibitem[{\citenamefont{Y.~Grossman}(1995)}]{Grossman:1995wx}
\bibinfo{author}{\bibfnamefont{Y.}~\bibnamefont{Grossman}}, 
  \bibinfo{journal}{Phys. Lett. B} \textbf{\bibinfo{volume}{359}}, \bibinfo{pages}{141}
  (\bibinfo{year}{1995}). 


    \bibitem[{\citenamefont{T.~Ohlsson}(2014)}]{Ohlsson:2013nna}
\bibinfo{author}{\bibfnamefont{T.}~\bibnamefont{Ohlsson}, \bibfnamefont{H.}~\bibnamefont{Zhang}
 and \bibfnamefont{S.}~\bibnamefont{Zhou},
  \bibinfo{journal}{Phys. Lett. B} \textbf{\bibinfo{volume}{728}}, \bibinfo{pages}{148}}
  (\bibinfo{year}{2014}). 
  

    \bibitem[{\citenamefont{D.~Meloni}(2009)}]{Meloni:2009cg}
\bibinfo{author}{\bibfnamefont{D.}~\bibnamefont{Meloni}, \bibfnamefont{T.}~\bibnamefont{Ohlsson},
 \bibfnamefont{W.}~\bibnamefont{Winter} and \bibfnamefont{H.}~\bibnamefont{Zhang},
  \bibinfo{journal}{JHEP} \textbf{\bibinfo{volume}{1004}}, \bibinfo{pages}{041}}
  (\bibinfo{year}{2010}). 


    \bibitem[{\citenamefont{J.~Koppw}(2008)}]{Kopp:2007ne}
\bibinfo{author}{\bibfnamefont{J.}~\bibnamefont{Kopp}, \bibfnamefont{M.}~\bibnamefont{Lindner}, 
\bibfnamefont{T.}~\bibnamefont{Ota} and \bibfnamefont{J.}~\bibnamefont{Sato}, 
  \bibinfo{journal}{Phys. Rev. D} \textbf{\bibinfo{volume}{77}}, \bibinfo{pages}{013007}}
  (\bibinfo{year}{2008}). 
  
  
     \bibitem[{\citenamefont{T.~Ohlsson}(2009)}]{Ohlsson:2008gx}
\bibinfo{author}{\bibfnamefont{T.}~\bibnamefont{Ohlsson} and \bibfnamefont{H.}~\bibnamefont{Zhang}, 
  \bibinfo{journal}{Phys. Lett. B} \textbf{\bibinfo{volume}{671}}, \bibinfo{pages}{99}}
  (\bibinfo{year}{2009}). 


   \bibitem[{\citenamefont{R.~Leitner}(2011)}]{Leitner:2011aa}
\bibinfo{author}{\bibfnamefont{R.}~\bibnamefont{Leitner}, \bibfnamefont{M.}~\bibnamefont{Malinsky},
 \bibfnamefont{B.}~\bibnamefont{Roskovec} and \bibfnamefont{H.}~\bibnamefont{Zhang}, 
  \bibinfo{journal}{JHEP} \textbf{\bibinfo{volume}{1112}}, \bibinfo{pages}{001}}
  (\bibinfo{year}{2011}). 

  
 \bibitem[{\citenamefont{C.~Biggio}(2009)}]{enrique}
\bibinfo{author}{\bibfnamefont{C.}~\bibnamefont{Biggio}, \bibfnamefont{M.}~\bibnamefont{Blennow}
and \bibfnamefont{E.}~\bibnamefont{Fernandez-Martinez}}, 
  \bibinfo{journal}{JHEP} \textbf{\bibinfo{volume}{0908}}, \bibinfo{pages}{090}
  (\bibinfo{year}{2009}). 
  
  
  \bibitem[{\citenamefont{T.~A.~Mueller}(2011)}]{Mueller:2011nm}
\bibinfo{author}{\bibfnamefont{T.}~\bibfnamefont{A.}~\bibnamefont{Mueller} \bibnamefont{et~al.},
  \bibinfo{journal}{Phys. Rev. C} \textbf{\bibinfo{volume}{83}}, \bibinfo{pages}{054615}}
  (\bibinfo{year}{2011}). 


  \bibitem[{\citenamefont{P.~Huber}(2011)}]{Huber:2011wv}
\bibinfo{author}{\bibfnamefont{P.}~\bibnamefont{Huber},
  \bibinfo{journal}{Phys. Rev. C} \textbf{\bibinfo{volume}{84}}, \bibinfo{pages}{024617}}
  (\bibinfo{year}{2011}); 
  \bibinfo{author}{Erratum-ibid.,
  \bibinfo{journal}{Phys. Rev. C} \textbf{\bibinfo{volume}{85}}, \bibinfo{pages}{029901}}
  (\bibinfo{year}{2012}).
  

 \bibitem[{\citenamefont{S.~Jetter}(2013)}]{talk:DayaBay}
\bibinfo{author}{\bibfnamefont{S.}~\bibnamefont{Jetter}, talk given at NuFact13}.


   \bibitem[{\citenamefont{F.~P.~An}(2014)}]{An:2013zwz}
\bibinfo{author}{\bibfnamefont{F.}~\bibfnamefont{P.}~\bibnamefont{An}} \bibnamefont{et~al.}
  (\bibinfo{collaboration}{Daya Bay Collaboration}), 
   \bibinfo{journal}{Phys. Rev. Lett.} \textbf{\bibinfo{volume}{112}}, \bibinfo{pages}{061801}
  (\bibinfo{year}{2014}). 
  
  

    \bibitem[{\citenamefont{P.~Vogel}(1999)}]{Vogel:1999zy}
\bibinfo{author}{\bibfnamefont{P.}~\bibnamefont{Vogel} and \bibfnamefont{J.}~\bibfnamefont{F.}~\bibnamefont{Beacom},
  \bibinfo{journal}{Phys. Rev. D} \textbf{\bibinfo{volume}{60}}, \bibinfo{pages}{053003}}
  (\bibinfo{year}{1999}). 
  
  
   \bibitem[{\citenamefont{P.~Huber}(2005)}]{GLOB2}
\bibinfo{author}{\bibfnamefont{P.}~\bibnamefont{Huber}, \bibfnamefont{J.}~\bibnamefont{Kopp},
\bibfnamefont{M.}~\bibnamefont{Lindner} and \bibfnamefont{W.}~\bibnamefont{Winter}, 
  \bibinfo{journal}{Comput. Phys. Commun.} \textbf{\bibinfo{volume}{167}}, \bibinfo{pages}{195}}
  (\bibinfo{year}{2005});
  \bibinfo{author}{\bibfnamefont{P.}~\bibnamefont{Huber}, \bibfnamefont{J.}~\bibnamefont{Kopp},
\bibfnamefont{M.}~\bibnamefont{Lindner}, \bibfnamefont{M.}~\bibnamefont{Rolinec} and \bibfnamefont{W.}~\bibnamefont{Winter}, 
  \bibinfo{journal}{Comput. Phys. Commun.} \textbf{\bibinfo{volume}{177}}, \bibinfo{pages}{432}}
  (\bibinfo{year}{2007}).


  \bibitem[{\citenamefont{F.~Capozzi}(2013)}]{Capozzi:2013csa}
\bibinfo{author}{\bibfnamefont{F.}~\bibnamefont{Capozzi}, \bibfnamefont{G.}~\bibfnamefont{L.}~\bibnamefont{Fogli}, \bibfnamefont{E.}~\bibnamefont{Lisi},
\bibfnamefont{A.}~\bibnamefont{Marrone}, \bibfnamefont{D.}~\bibnamefont{Montanino} and \bibfnamefont{A.}~\bibnamefont{Palazzo}},
 \eprint{arXiv:1312.2878}.
  

 \bibitem[{\citenamefont{K.~Eitel}(2005)}]{Eitel:2005hg}
\bibinfo{author}{\bibfnamefont{K.}~\bibnamefont{Eitel},
  \bibinfo{journal}{Nucl. Phys. Proc. Suppl.} \textbf{\bibinfo{volume}{143}}, \bibinfo{pages}{197}}
  (\bibinfo{year}{2005}). 
  

\end{thebibliography}
\end{document}